\documentclass[a4paper,12pt]{article}

\usepackage{amsmath}
\usepackage[final]{graphicx}
\usepackage{bm}
\usepackage{amssymb}

\def\epe{\varepsilon'/\varepsilon}

\def\eq#1{{eq.~(\ref{#1})}}

\def\Tr{\mathop{\mbox{Tr }}\,}

\def\ee{$\varepsilon'/\varepsilon$ }

\newcommand{\bea}{\begin{eqnarray}}
\newcommand{\beq}{\begin{equation}}
\newcommand{\eea}{\end{eqnarray}}
\newcommand{\eeq}{\end{equation}}
\newcommand{\nnu}{\nonumber}

\usepackage{amssymb}
\newcounter{comment}

{\refstepcounter{comment}%
\begin{quote}
\ttfamily\small$\blacksquare$ \textbf{\underline{Comment} $\sharp$\thecomment:}}%
{\end{quote}}

\begin{document}
\hfill
\begin{minipage}{20ex}\small
ZAGREB-ZTF-08-01\\
\end{minipage}

\begin{center}
\baselineskip=2\baselineskip
\textbf{\LARGE{
New Dipole Penguin Contribution \\ to 
$K\to \pi\pi$ decays
}}\\[6ex]
\baselineskip=0.5\baselineskip

{\large Jan~O.~Eeg$^{a,}$\footnote{j.o.eeg@fys.uio.no; corresponding author},  
Kre\v{s}imir~Kumeri\v{c}ki$^{b,}$\footnote{kkumer@phy.hr}, 
and
Ivica~Picek$^{b,}$\footnote{picek@phy.hr}}\\[4ex]
\begin{flushleft}
\it
$^{a}$Department of Physics, University of Oslo, P.O.B. 1048 Blindern, N-0316 Oslo, 
Norway\\[1.5ex]
$^{b}$Department of Physics, Faculty of Science, University of Zagreb,
 P.O.B. 331, HR-10002 Zagreb, Croatia\\[3ex]
\end{flushleft}
\today \\[5ex]
\end{center}

\begin{abstract}

We point out that the standard chromomagnetic penguin dipole operator
has a counterpart corresponding
to off-shell momenta for external quarks.
By employing the chiral quark model, we show that this new dipole penguin operator
has the same bosonisation as the standard $Q_6$ operator.
Accordingly,  this new operator enlarges by 
$\sim 5$~\%  the referent $Q_6$ contribution, which gives the 
dominant contribution
to the CP-violating ratio $\epe$
and also gives an important contribution to  the $\Delta I = 1/2$ amplitude.

\end{abstract}

\vspace*{2 ex}

\begin{flushleft}
\end{flushleft}

\clearpage

\section{Introduction}


The physics of the $K \rightarrow 2 \pi$ decay within the Standard Model has
been a great challenge. First, 
the $\Delta I = 1/2$ rule is only vaguely
understood, and similarly, the CP-violating quantity $\epe$
has been very difficult to estimate because of the inherent
hadronic uncertainties. 
In electroweak decays of K-mesons one
constructs an effective  Lagrangian at the quark level. Thereby one
uses the equations of motion. In QED it is known that the main part of the
Lamb shift disappears if the equation of motion is used for the
electron self-energy, warning us that it
is a bound state, nonperturbative effect. What about the
Lamb shift like effects in the QCD context of hadronic decays?

An early study by two of us \cite{ep93,ep94} was undertaken
 in order to account for
the off-shell effects in $K$-meson decays. 
In particular, in ref. \cite{ep93} we considered only the CP-conserving
 $K \rightarrow 2 \pi$ amplitude, and the
CP-violating off-shell part  has been assigned to  
``the waiting list of pieces to be included in the re-evaluation 
of $\epe$ ''. 
At roughly the same time Bertolini et al. \cite{BeFG94} 
included in such a reevaluation 
the chromomagnetic $Q_{11}$ penguin dipole contribution to $\epe$, and 
in the present Letter we find as appropriate to consider its off-shell counterpart.

{From} the very beginning there has been a close relation between the
 gluonic penguin operators and the  
attempts to predict the direct CP-violation parameter $\epe$ of 
the $K\rightarrow 2\pi$ amplitude. 
At first, the gluonic penguin had pointed to a possibility of a sizable (direct) CP-violating $K\rightarrow 2\pi$ amplitude.
However, there had been a turning point with
the large value of the t-quark mass, that led to a substantial cancellation between
the dominant (gluonic $Q_6$, and electroweak $Q_8$) penguin contributions. 
That called for
investigation of other possible contributions, including the off-shell
contribution at hand.

For a brief history of progress on evaluating $\epe$ and a more complete
 list of references on evaluation of $\epe$ we refer to the review 
in ref. \cite{BeFE00}, and for later work to ref. \cite{epslater}.
These references show the way in which $\epe$ is structured over the 
contributions from the operators belonging to, by now, standard operator
 basis \cite{BuBL96}. The relatively large value for $\epe$ obtained by the
 Trieste-Oslo collaboration \cite{Bertolini:1997nf} turned out to be a successful
 prediction for the outcome of the subsequent
 Fermilab \cite{KTeV} and CERN \cite{NA48} measurements, that together
 with previous results from NA31 \cite{Barr:1993rx} and E731 \cite{Gibbons:1993zq}
 experiments reported the world average
\begin{equation}\label{eps}
\epe=(16.7\pm 1.6) \cdot 10^{-4}.
\end{equation}
Basically, the prediction by the Trieste-Oslo group resides in the
use of the chiral quark model ($\chi$QM) \cite{Weinb/Manohar,Biea,PdeR}
which enabled to account for unavoidable nonperturbative QCD effects. 
Despite being a model approach, such a treatment has some unique features:
Besides weakening the destructive interference between the matrix elements of
$Q_6$ and $Q_8$ operators (which we explicate in the next section), it enables
the evaluation of the matrix elements for all the relevant operators within a single framework.
Let us stress that in the chiral quark model we could evaluate the mentioned 
off-shell effects in K-meson decays \cite{ep93,ep94,k3p95} and to
account for the off-shell, off-diagonal self energy contribution to
the CP-violating ratio $\epe$ \cite{Bergan:1996sb}.
We believe that the off-shell chromomagnetic effect, calculated in this
Letter, presents an interesting new piece in illuminating the $K\to \pi \pi$ puzzle.

\section{Effective Lagrangian for $\Delta S =1$  Decays}

The amplitudes for
$K \rightarrow 2 \pi$
are described  by an effective weak  Lagrangian at quark 
level \cite{BuBL96,Roma}
 \beq
 {\cal L}_{W}=  \sum_i  C_i(\mu) \; Q_i (\mu) \; ,
 \label{Lquark}
\eeq
where all information on the short distance (SD) loop 
effects above a renormalisation scale $\mu$
is contained in Wilson coefficients $C_i$.
These depend on the masses of the $W,Z$-bosons,
  the heavy quark masses ($m_q > \mu$), 
 $\Lambda_{QCD}$ and on the 
renormalisation scheme.
The $Q_i$'s are quark operators,
typically containing products of two quark currents.

The standard
basis (for $\mu < m_c$, relevant to kaon decays) includes  ten  4-quark operators.
 We display the four most
important for the $\Delta I = 1/2$ rule ($Q_{1,2,6}$)
and \ee ($Q_{6,8}$):
\begin{eqnarray}
Q_{1}  =  4 \,  ( \overline{s}_L \gamma^\mu  d_L )  \; \,
           ( \overline{u}_L \gamma_\mu  u_L )
\; \;  , & \,
Q_{6}  = -8  \sum_{q} \, ( \overline{s}_L q_R ) \; \, ( \overline{q}_R  d_L )
\, , \nnu  \\
Q_{2}  =  4 \, ( \overline{s}_L \gamma^\mu u_L ) \; \,
           ( \overline{u}_L \gamma_\mu d_L )
\; \; , & \,
Q_{8}  =  -12 \sum_{q} \hat{e}_q \; ( \overline{s}_L q_R )                                       \; \, ( \overline{q}_R  d_L )
\, , 
\label{Q1-10} 
\end{eqnarray}  
where  $\hat{e}_q$  are the quark charges 
($\hat{e}_u = 2/3$,  $\hat{e}_d=\hat{e}_s =-1/3$), and $q_{L,R}$
are the left- and right-handed projections of the quark fields.

Some studies also include the standard  chromomagnetic dipole
operator \cite{BeFG94,BeFE00,BuBL96,BEFQ11} that
 can be written as 
\beq
 Q_{11} \, = \, \frac{g_s}{8 \pi^2}
 \overline{d} \; \Bigl[ m_s \, R + m_d \, L \Bigr]
 \,  \, \sigma  \cdot  G
 \;  s  \,  + \, h.c.  \; ,
\label{11}
\eeq
where $ \sigma  \cdot \, G  \equiv
\sigma^{\mu \nu}\, G^a_{\mu \nu} t^a$,
 $G^a_{\mu \nu}$ is the gluon field tensor, $t^a$ are the 
SU(3) generators normalised as $\Tr( t^a t^b ) = \delta^{ab}/2$,
 and $(L,R)=(1\mp\gamma_5)/2$.
However, we should stress that the operator for  $s \to d + gluon(s)$ 
transition generated by loop diagrams \cite{ep93,SiW90} is not $Q_{11}$ in eq. (\ref{11}), but 
is given by 
\begin{eqnarray}
{\cal{L}}(s\rightarrow dG) \, & = & \,  B \, 
\epsilon^{\mu \nu \lambda \rho} \,
G^a_{\mu \nu} \, ( \bar{d}_{L} \;  t^a \, i \stackrel{\leftrightarrow}{D_{\lambda}}
 \gamma_{\rho} s_L )   \nonumber \\
 & \rightarrow & \, - \, \frac{1}{2} \, B  \, \bar{d_L} [i\gamma \cdot D  \,
 \sigma \cdot G  +  \sigma \cdot G \, i\gamma \cdot D ] \, s_L \; .
\label{eq:lagg}
\end{eqnarray}
Here the second line is obtained by using simple algebra of Dirac matrices, and the coefficient
 $B \sim g_s G_F \lambda_{KM}$  depends on the loop integration.
 It is convenient 
to rewrite (\ref{eq:lagg}) 
as a sum of an off-shell term  ${\cal{L}}_G$  and
the chromomagnetic moment term ${\cal{L}}_{\sigma}$ \cite{ep93} :
\begin{eqnarray}
{\cal{L}}(s\rightarrow d G) \,  = \, {\cal{L}}_G \, + 
{\cal{L}}_{\sigma} \; ; 
\nonumber \\
{\cal{L}}_G  =  C_G \; Q_G \quad  ;  \; 
{\cal{L}}_{\sigma}  =  C_{11} \; Q_{11} \; ,
\label{eq:lags}
\end{eqnarray}
where we have introduced 
a counterpart  of the
standard dipole operator in eq. (\ref{11})~:
\beq
 Q_{G} \, = \, \frac{g_s}{8 \pi^2}
 \overline{d} \; \Bigl[(i\gamma \cdot D - m_d) \,
 \sigma \cdot G \, L 
\, + \, \sigma \cdot G \, R (i\gamma
\cdot D -m_s)  \Bigr]
s  \,  + \, h.c.  \, .
\label{Q_G}
\eeq
This operator vanishes by the QCD equation of motion 
for perturbatively interacting quark fields.
The coefficients $C_G$ and $C_{11}$ above,
being equal at the $W$-scale, evolve differently down to 
the scale $\sim$ 1 GeV, where hadronic matrix elements are evaluated.
 Therefore, 
in the next section we consider SD QCD corrections 
to the Wilson coefficient $C_G$, that to our knowledge are not 
 given in the literature.

In order to
keep and calculate the contributions from $Q_G$,
 one needs a framework to
incorporate the effects of off-shell quarks at low energies, 
or equivalently,  a framework where the operator $Q_G$ cannot be rotated away.
An important point of this Letter is that the operator 
 $Q_G$, to leading order, has the same
 bosonisation as the $Q_6$ operator. 
This fact will enable us a direct comparison of the
off-shell contribution coming from the operator $Q_G$, and the leading
CP-violating and CP-conserving
contributions stemming from $Q_6$.

The Lagrangian (\ref{eq:lags}) is just a part of the more complete effective
 weak Lagrangian at quark level relevant to $K$-decays
\begin{equation}
 {\cal{L}}_W \; = \; {\cal{L}}_{4q} \, + \, 
 {\cal{L}}_\sigma \, + \, 
{\cal{L}}_G \,  +  \,
 {\cal{L}}_{sd}^R \; ,
\label{LW}
\end{equation}
where the additional terms are ${\cal{L}}_{4q}$ corresponding to standard four
quark operators (\ref{Q1-10}) and  ${\cal{L}}_{sd}^R$, the renormalised
 off-diagonal self-energy \cite{ep93,SiW90}
\begin{equation}
 {\cal{L}}_{sd}^R \; = \;
  -  A \, \bar{d}(i\gamma \cdot D - m_d) (i \gamma \cdot D R + M_R R + M_L L) 
(i\gamma \cdot D - m_s) s \ . 
\end{equation}
The most important part of this last term for
 $\epe$, the so-called self-penguin, was critically examined
in ref. \cite{Guberina:1986cb} and was shown to have
a considerable off-shell contribution in  $\chi$QM  framework of ref.
\cite{Bergan:1996sb}.

 In the standard SD procedure \cite{BuBL96,Roma},
${\cal{L}}_G$ and ${\cal{L}}_{ds}^R$ in ${\cal{L}}_W$,
eq. (\ref{LW}), would be absent when
applying the equation of motion at quark level.
Instead, a 
more appropriate procedure should be to
 transform these terms away  by a field redefinition, introducing
 new quark fields
\begin{eqnarray}
d' = d + B \sigma \cdot G L s 
- \frac{1}{2} A (i \gamma \cdot D R + M_R R + M_L L) 
(i\gamma \cdot D - m_s) s \ , 
  \nonumber \\
s' = s + B^* \sigma \cdot G L d 
- \frac{1}{2} A^* (i \gamma \cdot D R + M_R L + M_L R) 
(i\gamma \cdot D - m_s) d \ .
\label{Fred}
\end{eqnarray}
Then the parts  ${\cal{L}}_G$ and ${\cal{L}}_{ds}^R$ 
 involving the covariant derivatives are
apparently removed,
absorbed in the Dirac Lagrangian ${\cal{L}}_{f}(q)$ for $q=(u,d,s)$
 \cite{ep93}:
\begin{equation}
 {\cal{L}}_{f}(q) + {\cal{L}}_G +
{\cal{L}}_{ds}^R \, = \, {\cal{L}}_{f}(q') \; ,
\end{equation}
where ${\cal{L}}_{f}(q')$ is given later in eq. (\ref{QCDf}), with $q$ replaced 
by $q'$. 
In a strict  SD treatment, primed and unprimed quark fields are equivalent.
This means that  ${\cal{L}}_G$ (and similarly ${\cal{L}}_{ds}^R$)
does not contribute to $s \rightarrow d$ transitions
 for on-shell external quarks.
In section 4 we will show how the effects of the mentioned off-shell operators reappear 
when low-energy strong interactions are taken into account in terms
of the $\chi$QM.

\section{The Wilson coefficient $C_G$}

It is convenient to distinguish the CP-conserving and CP-violating parts of the
Wilson coefficients for the  $\Delta S=1$ quark operators
  in (\ref{Lquark}).
At some scale $\mu$ they can be written as
\bea
C_i(\mu) =  - \frac{G_F}{\sqrt{2}}  
 \Bigl[ \lambda_u \, z_i(\mu) - \lambda_t \,  y_i(\mu) \Bigr] \; ,
\label{Lqcoef}
\eea
where $G_F$ is the Fermi coupling, the functions $z_i(\mu)$ and
 $y_i(\mu)$ are the CP-conserving and CP-violating parts of the
 coefficients, respectively, and  
$\lambda_q = V_{qd}\,V^*_{qs}$ (for $q=u,t$) are the CKM factors.
The numerical values of $z_i$ and $y_i$ 
are in the range of order one down to $10^{-4}$, and can be found in the
literature \cite{BuBL96} for operators up to $Q_{11}$.
In what follows we calculate the corresponding values for the operator
$Q_G$, restricted to the truncated basis given
 by $Q_\pm = (Q_2 \pm Q_1)/2$ and $Q_G$ operators.
Thereby we denote by $a_{\pm}$ the Wilson coefficients
 of the 4-quark operators $Q_{\pm}$ with diagonal
 anomalous dimension matrix.
In the logarithmic approximation, for $\mu \le m_c$ they are 
\begin{equation}
   a_{\pm}(\mu^{2})=
 \left[\frac{\alpha_{s}(m_c^{2})}{\alpha_{s}(\mu^2)}
      \right]^{\frac{d_{\pm}}{b(3)}} 
 \left[\frac{\alpha_{s}(m_b^{2})}{\alpha_{s}(m_c^{2})}
      \right]^{\frac{d_{\pm}}{b(4)}} 
 \left[\frac{\alpha_{s}(M_W^{2})}{\alpha_{s}(m_b^{2})}
      \right]^{\frac{d_{\pm}}{b(5)}} 
    \quad,
\label{cQCD}
\end{equation}
where  $d_{+}=+2$ and $d_{-}=-4$ are the anomalous dimensions
and $b(N_f) = 11 N_c/3 - 2 N_f/3$, where
$N_f$ is the number of active flavours.
The coefficient  $a_{\pm}$ in the equation above, will become either
the function $z(\mu)$ for the CP conserving part, or the function
$y(\mu)$ for the CP violating part. For these cases, 
the values of $\mu$ will be
$\mu\simeq 1$ GeV or $\mu=m_c$, respectively.

The anomalous dimension matrix for the truncated basis of the three operators 
$Q_\pm$ and $Q_n$ has, for 
$n= 6,11,G$  the form:
\begin{eqnarray}
  \gamma =
 \left[\begin{array}{ccc} d_+ & 0
&X_+ \\ 0 &  d_-  & X_-  \\
0 & 0 &  Y_n  \end{array}\right] \; .
\label{adm}
\end{eqnarray}
where
\begin{align}
X_\pm& = \begin{cases}
 \frac{11 N_c}{18} - \frac{29}{18 N_c} \pm \frac{3}{2} & \text{for $n=11,G$}\\
 \frac{1}{6} & \text{for $n=6$}
 \end{cases}  \\
Y_n& = \begin{cases}
-6 \left(\frac{N_{c}^2-1}{2N_c}\right) + \frac{N_f}{3} & \text{for n=6} \\
-2 N_c - \frac{4}{N_c} & \text{for n=11} \\
0 & \text{for n=G} 
\end{cases}
\end{align}
One should note that  $Q_G$, being an 
operator that vanishes by the perturbative QCD equation of motion, has zero anomalous dimension,
 $Y_G=0$, in contrast to  $Q_6$ and $Q_{11}$.

For handling  the leading QCD corrections, 
there is 
a suitable prescription introduced in Refs. \cite{NSVZ,SVZemp}
and applied by others \cite{Oth,PRDjoe,ENP90,EKP96}.  
Using this prescription, one can write the amplitude
as an integral over virtual quark loop momenta. 
The QCD-corrected coefficients $z_G$ and $y_G$ can be
expressed in the integral form, which for CP-conserving case reads
\begin{equation}
z_G=  \, \int_{\mu^2}^{m_{c}^{2}}
\frac{dp^{2}}{p^{2}}  \left( \frac{\alpha_s(p^{2})}{4 \pi} \right)
\left[ X_+ \, a_+(p^2) \, +  X_- \, a_-(p^2) \right]/2 
  \; .
\label{aDQCD}
\end{equation}
For the  CP-violating case we have 
\begin{equation}
y_G= (F_2^c - F_2^t) + \, \int_{m_{c}^2}^{M_{W}^{2}}
\frac{dp^{2}}{p^{2}}  \left( \frac{\alpha_s(p^{2})}{4 \pi} \right)
\left[ X_+ \, a_+(p^2) \, +  X_- \, a_-(p^2) \right]/2 
  \; ,
\end{equation}
where $F_{2}^{i}\equiv F_{2}(m_{i}^2/M_{W}^2)$ are well-known Inami-Lim
functions \cite{InL}.
Similarly, repeating the standard renormalisation group procedure of ref.
\cite{BuBL96,GSW}, gives for the  Wilson coefficient of $Q_G$, for the
CP-conserving case, 
\beq
z_G(\mu) = \frac{X_+}{d_+} \left(\eta^{\hat{d}_+} -1\right) a_+(m_c)
               + \frac{X_-}{d_-}\left(\eta^{\hat{d}_-} -1\right)a_-(m_c)
              \; ,
\label{eq:rgs}
\eeq
where  $\eta$ = $\alpha_s(m_c)/\alpha_s(\mu)$, and $\hat{d}_\pm \equiv
d_\pm/b(3)$. For the coefficient relevant for CP violation one obtains
\beq
y_{G}(\mu = m_c) = (F_2^c - F_2^t) 
              + \frac{X_+}{ d_+} \left(a_+(m_c) -1\right)
               + \frac{X_-}{ d_-} \left(a_-(m_c) -1\right) \; .
\label{eq:rgf}
\eeq
These expressions lead us to values for Wilson coefficient of $Q_G$ displayed in 
Table 1. In this Table
we also give our values for coefficients of $Q_6$ and  $Q_{11}$ that 
conform to those given in \cite{BuBL96}.
\begin{table}
\centering
\begin{tabular}[ht]{cccc}\hline
  & $Q_6$  & $Q_{11}$ & $Q_G$ \\ \hline
$z(\mu = 1\,{\rm GeV})$ & -0.009 &  -0.033 & -0.035 \\
$y(\mu = m_c )$ & -0.083 & -0.318 & -0.407  \\ \hline
\end{tabular}
\caption{Wilson coefficients of relevant operators obtained by
leading order renormalisation group evolution from scale
$\mu = M_W$, with flavour number thresholds at $m_b = 4.4$ GeV and
$m_c = 1.3$ GeV, using $\alpha_{\rm s}(M_Z) = 0.117$.}
\label{tab:yz}
\end{table}

\section{Bosonisation in the Chiral Quark Model}

For  light pseudoscalar mesons there is
a well defined effective theory, chiral perturbation theory ($\chi PT$),
having the basic symmetries of QCD.
One  can try to match $\chi PT$ to the weak Lagrangian at quark level, eq. (\ref{Lquark}),
 by  {\em bosonizing} the quark 
operators $Q_i$:
\beq
 Q_i \rightarrow  \sum_j  F_{ij} \; \hat{{\cal L}}_j \; \; ,
 \label{BosQ}
\eeq
where the $\hat{{\cal L}}_j$'s
are chiral Lagrangian terms having the symmetry of $Q_i$, and  $F_{ij}$ are quantities
  to be calculated with 
non-perturbative methods (including quark models).
Knowing the bosonization in (\ref{BosQ}),
 we could calculate the various $K$-decay amplitudes
from a $\Delta S = 1$ chiral Lagrangian
\beq
 {\cal L}_{W}(\chi PT) =  \sum_j   G_j \; \hat{{\cal L}}_j \;  ; \qquad
 G_j =  \sum_i C_i \;  F_{ij} \; .
 \label{LMes}
\eeq
The idea of such an approach is that the coefficients should be 
calculated (and matched) at the border  of
the SD and LD regimes.
Thereby the  factorized form of the coefficients $G_j$ in eq. (\ref{LMes})
 explicates the separation of SD contributions sitting in the $C_i$'s, and LD 
contributions residing in the $F_{ij}$'s.

In order to bosonise our relevant operators we employ the 
chiral quark model ($\chi$QM), that has been advocated by 
 many authors \cite{Weinb/Manohar,Biea,PdeR}
as an  effective low-energy QCD.
In this model, chiral-symmetry breaking is
taken into account by adding a term to ordinary QCD:
\begin{equation}
{\cal{L}}_{QCD} \rightarrow {\cal{L}}_{QCD\chi} = {\cal{L}}_{QCD}
+  {\cal{L}}_{\chi} \; , 
\label{QCDchi}
\end{equation}
where ${\cal{L}}_{QCD}$ in addition to the pure gluonic part contains
the fermionic part
\begin{eqnarray}
 {\cal{L}}_{f}(q) = \bar{q} (i \gamma \cdot D - {\cal{M}}_q) q \; \;
; \; \; \;
q  \; = \;
 \left( \begin{array}{c} u \\  d\\  s \end{array}
\right)  \; \; .
\label{QCDf}
\end{eqnarray}
Here ${\cal{M}}_q= \,\mbox{diag} \,(m_u,m_d,m_s)$ 
is the {\em current} quark mass-matrix, whereas 
a non-perturbative term in (\ref{QCDchi}), 
\begin{equation}
  {\cal{L}}_{\chi} = - m (\overline{q_L} \, \Sigma^\dagger \, q_R + 
\overline{q_R} \, \Sigma \, q_L) \; ,
\label{LChiQM}
\end{equation}
contains the parameter  $m$ that is interpreted as the
{\em constituent} quark mass ($\sim 200-250$ MeV).
Note that the constituent and current masses are tied to
different terms in the Lagrangian, with different transformation properties.
Here $q$ is the SU(3) flavour triplet quark field, and $\Sigma$ contains
 the Goldstone-octet fields $\pi^a$:
\begin{equation}
\Sigma = exp(i\sum_a \lambda^a \pi^a / f_{\pi}) \; \; ,
\end{equation}
where $\lambda^a$
are the Gell-Mann matrices.
 The term ${\cal{L}}_{\chi}$ contains  meson-quark couplings.
This means that the quarks can be integrated out and the coefficients
of the various terms in  the chiral Lagrangian are calculable
 from ${\cal{L}}_{QCD\chi}$ in eq. (\ref{QCDchi}).

Our procedure
 based on the equations (\ref{QCDchi}, \ref{QCDf}, \ref{LChiQM})
 has to be understood in
the following way: At scales  above the cut-off $\Lambda_{\chi}$, the
total Lagrangian is the sum of the standard ${\cal{L}}_{QCD}$ and the weak
effective Lagrangian  ${\cal{L}}_W$ in eq. (\ref{LW}), given in terms of 
 the physical $u,d,s$-fields. Then at
scales below $\Lambda_{\chi}$, the term  ${\cal{L}}_{\chi}$ in
 eq. (\ref{LChiQM}) is
turned on, so that the matrix elements of  ${\cal{L}}_W$ between
mesonic states can be calculated owing to the meson-quark couplings.

The model has a  ``rotated''  picture, where
 the term 
${\cal{L}}_{\chi}$ 
in (\ref{LChiQM}) is transformed into a pure
mass term $- m \overline{\chi} {\chi}$
for flavour rotated ``constituent quark'' fields ${\chi}_{L,R}$ :
\beq
q_L \rightarrow  {\chi}_L =  \xi  q_L \quad \mbox{and} \quad
q_R \rightarrow  {\chi}_R =  \xi^\dagger q_R \, ,
\label{Flavrot}
\eeq
where $\xi \, \cdot \, \xi  = \Sigma $.
The meson--quark couplings in this rotated  picture  arise from
the kinetic
(Dirac) part of the constituent quark Lagrangian.
These interactions can be described in terms of vector and axial vector
fields coupled to constituent quark fields
${\chi} = {\chi}_R + {\chi}_L$.
The sum of ${\cal L}_f$ in (\ref{QCDf}) and ${\cal L}_\chi$ in (\ref{LChiQM})
are transformed into the equivalent form
\begin{equation}
{\cal L}_{\chi QM} =  
\overline{{\chi}} \Big[\gamma^\mu (i D_\mu   +    {\cal V}_{\mu}  +  
\gamma_5  {\cal A}_{\mu})    -    m \Big]\chi 
  -    \overline{{\chi}} \widetilde{M_q} \chi \;  , 
\label{chqmR}
\end{equation}
where
\beq
{\cal{V}}_{\mu} \, = \, \frac{1}{2} \left[\xi^\dagger (i \partial_{\mu} \xi) 
\, + \,   \xi \, (i \partial_{\mu} \, \xi^\dagger)\right]\; \; ,
\quad
{\cal{A}}_{\mu} \, = \, \frac{1}{2}  \left[\xi^\dagger (i \partial_{\mu} \xi) 
\, - \,   \xi \, (i \partial_{\mu} \, \xi^\dagger)\right] \; \, ,
\label{vectors}
\eeq
and the current quark masses are residing in
\bea
\widetilde{M_q} = \widetilde{{M}_q}^V + \widetilde{{M}_q}^A \gamma_5 
\; , \quad \mbox{with} \quad
 \; 
\widetilde{{M}_q}^{V(A)} \, = \, 
  \frac{1}{2}(\xi^\dagger {\cal M}_q \xi^\dagger \,
\pm \xi {\cal M}_q^\dagger \xi ) \;.
\label{curmass}
\eea

Now, having quarks that are exposed to strong interactions, and are
 described by the chiral quark model at hand, we have to
use the inverse of the transformation of eq. (\ref{Fred}) 
into eq. (\ref{LChiQM}). In this way the apparently removed terms
 are effectively reappearing 
  as a new term in ${\cal{L}}_{\chi}$ :
\begin{equation}
{\cal{L}}_{\chi}(q) = {\cal{L}}_{\chi}(q') + \Delta {\cal{L}}_{\chi}(q')
+ O(G_F^2) \; ,
\end{equation}
where the new term $\Delta {\cal{L}}_{\chi}(q') \sim G_F$ introduces new
vertices which compensate for those in ${\cal{L}}_G$ and
${\cal{L}}_{ds}^R$. When the fields $q'$ are integrated out,
 the result for
a physical amplitude at  mesonic level will be the same as if
${\cal{L}}_G$ and
${\cal{L}}_{ds}^R$ were applied without the field redefinition.

In general, each term in ${\cal{L}}_W$ which contains at least one
power of $(i \gamma \cdot D - m_q)$ may be removed
by a transformation like (\ref{Fred}). However, for each term which is
removed from ${\cal{L}}_W$, there will be a corresponding term
appearing in ${\cal{L}}_{\chi}$.
For the dipole operator $Q_G$, we obtain a contribution
proportional to the constituent mass $m$
\bea
 \Delta {\cal{L}}_{\chi}(q')_G \, =  \, \frac{C_G}{8 \pi^2} \, m \, 
 \left( \overline{q'}_L \, 
\lambda_- \, \sigma \,
 \cdot G \, \Sigma \, q'_R \, + \, 
 \overline{q'}_R \, \Sigma^\dagger 
 \sigma \,
 \cdot G  \, \lambda_-^\dagger \, q'_L \right) \; ,
\label{DipUnrot}
\eea
where $\lambda_-  =  (\lambda_6 - i \, \lambda_7)/2$ 
is the combination of Gell-Mann matrices
 which transforms an $s$-quark into a $d$-quark. Employing the flavour 
rotation from eq. (\ref{Flavrot}) we obtain the simple expression
\bea
 \Delta {\cal{L}}_{\chi}(q')_G \, =  \, \frac{C_G}{8 \pi^2} \, m \, 
 \overline{\chi'}  \, 
F_{(-)} \sigma \,
 \cdot G  \chi' \, + \, h.c. \; ,
\label{DFR}
\eea
where
\begin{equation}
F_{(-)} \; = \,  \xi \, \lambda_- \xi^{\dagger} \; .
\label{Fminus}
\end{equation}
The expression (\ref{DFR}) is ideal for bosonisation in terms of quark 
loops, whereas the analogous term for the self-energy still 
contains two derivatives, and 
can be  calculated in a different way, as done in ref \cite{Bergan:1996sb}.

\vspace*{2ex}
Using (\ref{chqmR}), the  strong chiral Lagrangian  $O (p^2)$
 can be understood
as two axial currents ${\cal{A}}_{\mu}$ attached to a quark loop, leading
to
\begin{equation}
{\cal{L}}^{(2)}_{s} \, \sim \,  \Tr \Bigl[ {\cal{A}}_{\mu} \, {\cal{A}}^{\mu}
 \Bigr] \, .
\end{equation}
Using the relations 
\begin{equation}
   2 i \, {\cal{A}}_{\mu} \, = \; - \,
\xi^{\dagger} (D^{\mu} \Sigma) \xi^{\dagger} \; = \;
 \xi (D_{\mu} \Sigma^{\dagger}) \xi \, , \label{9}
\end{equation}
one obtains the leading strong chiral Lagrangian
\begin{equation}
{\cal{L}}^{(2)}_{s} \, = \, \frac{f^2}{4} \,
 \Tr \Bigl( D^{\mu} \Sigma^{\dagger} \, D_{\mu} \Sigma \Bigr) \; ,
\end{equation}
where $D_\mu$ is the covariant derivative.
Note  that
${\cal{A}}_\mu$ is invariant under
local chiral transformations~\cite{Weinb/Manohar,Biea},
 in agreement with the invariance of
 ${\cal{L}}^{(2)}_s$. In contrast,
 the vector field ${\cal{V}}_\mu$ transforms as a gauge field.
Attaching in addition to  two ${\cal{A}}_\mu$'s also the mass term 
structures in (\ref{curmass}), we will obtain the well-known $L_5$ 
term which enters the matrix element of $Q_6$.

In addition to the $Q_6$ operator,
the referent object to which we compare our  
new off-shell dipole  penguin is the chromomagnetic dipole operator \eq{11}.  
It can be written in a
chiral  $SU(3)$  invariant form, as a first step in its bosonisation procedure: 
\begin{equation}
 Q_{11} \, = \,   \frac{g_s}{8 \pi^2} \, \Bigl[
 \overline{q}_R \; {\cal{M}}_{q} \lambda_-
 \; \sigma  \cdot  G \, q_L \, + \,
\overline{q}_L \;
 \; \sigma  \cdot  G \;
\lambda_- {\cal{M}}_{q}^{\dagger}  \;  q_R \Bigr] \, , \label{12}
\end{equation}
 where $q  = (u, d, s)$.
 Note that this
 operator transforms as $(\underline{8}_L, \underline{1}_R)$ under
the chiral $SU(3)_L \times SU(3)_R$ symmetry if
the current quark matrix is taken to transform as
${\cal{M}}_{q}\to V_R{\cal{M}}_{q}V_L^\dagger$, where $V_R$ and $V_L$ are the
chiral $SU(3)$ transformation matrices.

In the next step, we write $Q_{11}$
in the flavour rotated picture:
\beq
Q_{11} \, = \, \frac{g_s}{8 \pi^2} \, \overline{\chi} \Bigl[ F_{(-)}^V \, +
 \, F_{(-)}^A \, \gamma_5 \Bigr] \:
\sigma  \cdot  G \; \chi \, , \label{11r}
\eeq
where $F_{(-)}^{V,A} = \left(F_{(-)}^R \pm F_{(-)}^L\right) /2$ are expressed
in terms of
\beq
 F_{(-)}^L \; = \,  \xi^{\dagger} \, {\cal{M}}_{q}
         \lambda_- \xi^{\dagger}  \, , \qquad \mbox{and} \qquad
 F_{(-)}^R \; = \; \xi \, \lambda_-
{\cal{M}}_{q}^{\dagger} \,  \xi \, . \label{13}
\eeq

\begin{figure}
\centerline{\includegraphics[scale=0.8]{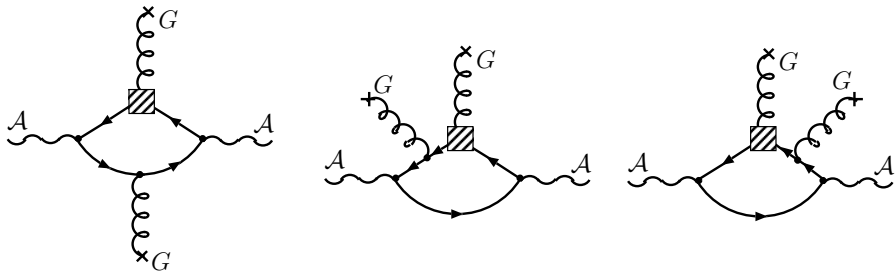}}
\caption{Bosonisation of the  operators $Q_{11}$ and $Q_{G}$,
where shaded squares represent appropriate insertions
from eqs. (\ref{11r}) and (\ref{DFR}), respectively}
\label{fig:BosonFRP}
\end{figure}
This operator is understood in terms of a quark loop.
Let us first stress that its lowest-order 
contribution is $\Tr(F_{(-)}^V)$, which is cancelled 
according to the FKW theorem~\cite{FKW}.
To the NLO, we obtain a term corresponding to an interaction
of $F_{(-)}^V$ and two axial fields attached to a quark loop, shown 
in Fig.~\ref{fig:BosonFRP}:
\begin{equation}
 {\cal{L}}^{(4)}(Q_{11}) \, \sim \,
\Tr \Bigl[ F_{(-)}^V  {\cal{A}}_{\mu} \, {\cal{A}}^{\mu} \Bigr]
  \,  . \label{15}
\end{equation}
Note that $F_{(-)}^A$ is not contributing, 
because there must be an even number of
 $\gamma_5$'s in the quark loop.
 Using (\ref{9}) and (\ref{13}), we find that
${\cal{L}}^{(4)}(Q_{11})$ can be written in the
final bosonised form
\begin{equation}
 {\cal{L}}^{(4)}(Q_{11})    =
G_{\underline{8}}^{(4)} (Q_{11}) \, \Tr  \Bigl[\left( 
\Sigma^{\dagger} \, {\cal{M}}_{q} \lambda_-
                 +   \lambda_- {\cal{M}}_{q}^{\dagger}\Sigma\right)
  \, D^{\mu} \Sigma^{\dagger} D_{\mu} \Sigma  \Bigr]  \, . 
\label{16}
\end{equation}
The coefficient $G_{\underline{8}}^{(4)} (Q_{11})$
was calculated in \cite{BEFQ11} to be 
$\, \sim  \, \langle \frac{\alpha_s}{\pi}G^2 \rangle \, C_{11}/(8 \pi^2) $,  
where the  two-gluon condensate is a model dependent quantity in our
approach. Finally,
one can deduce \cite{BEFQ11} the $K \rightarrow 2 \pi$ amplitude from the chiral structure
of ${\cal{L}}^{(4)}(Q_{11})$:
\begin{equation}
{\cal A}(K^0 \rightarrow \pi^+ \pi^- ; Q_{11}) \, = \,
\frac{\sqrt{2}}{f^3} \, (m_s - m_d)
 \, m_{\pi}^2 \,\, G_{\underline{8}}^{(4)} (Q_{11}) \; .
 \label{2.13} 
\end{equation}

Even though the coefficient in ${\cal{L}}^{(4)}(Q_{11})$ is large,
the $Q_{11}$ contribution to $K \rightarrow 2\pi$ is small because of the 
factor $m_{\pi}^2$ (in place of $m_K^2$ obtained for $Q_6$). 
Therefore, the modest role played
by $Q_{11}$ in $\varepsilon '/\varepsilon$ 
is due to this kinematical
suppression rather than due to being NLO in the chiral 
expansion.

The bosonisation of $Q_6$ follows basically the same line as for
$Q_{11}$ above.
The standard expression for $Q_6$ 
obtained by Fierz transformation and displayed in (\ref{Q1-10})
 can be rewritten  in the rotated picture as
\beq
Q_6 \, = \, - \, 8 \, (F_{(-)})_{\alpha \beta}
 \, (\overline{\chi}_L)_{\alpha} (\chi_R)_{\delta}
(\overline{\chi}_R)_{\delta} (\chi_L)_{\beta} \, ,
\eeq
where $F_{(-)} \; = \,  \xi \, \lambda_- \xi^{\dagger}$,
and Greek letters represent the flavour indices.
 Thus, likewise to ${\cal{L}}^{(4)}(Q_{11})$, the chiral representation
of $Q_6$ to leading order can be written as
\begin{equation}
 {\cal{L}}^{(2)}(Q_{6}) \, \sim \,
\Tr \Bigl [F_{(-)}  {\cal{A}}_{\mu} \, {\cal{A}}^{\mu} \Bigr]  \, . 
\label{23}
\end{equation}
By means of \eq{9} this can be written in the same
form as other familiar 
$\Delta S = 1$  octet operators  $O (p^2)$~\cite{cro67}
\begin{equation}
{\cal{L}}^{(2)}(Q_6) \, = \, G^{(2)}_{\underline{8}}(Q_6) \,
 \Tr \Bigl( \lambda_- D^{\mu} \Sigma^{\dagger} D_{\mu}
 \Sigma \Bigr) \, .
\label{L2(Q6)}
\end{equation}
This term  gives rise to the $K \rightarrow 2 \pi$ amplitude
\bea
{\cal A}( K^0 \rightarrow \pi^+ \pi^- ;\, Q_6) \, = \,
\frac{\sqrt{2}}{f^3}  \Bigl[ m_K^2 - m_\pi^2
\Bigr]\, G_{\underline{8}}^{(2)}(Q_6) \, .
\label{A2(Q6)}
\eea
The relevant coefficient in this expression has been calculated 
\cite{BeFE00} to be
\bea
 G_{\underline{8}}^{(2)}(Q_6) 
 \; = \; -16 \, C_6 \, \frac{L_5}{f_\pi^2} |\langle
\bar{q} q  \rangle |^2  \, ,
\label{G2(Q6)}
\eea
where $L_5 \simeq 1.4 \times 10^{-3}$
is the coefficient of the ${\cal O}(p^4)$ chiral strong Lagrangian 
mentioned at the beginning of this 
section \cite{BeFE00,Bertolini:1997nf,BEFQ11}.

Finally, as indicated in eq. (\ref{DFR}), the bosonisation of the operator
  $Q_G$ proceeds by inserting the expression (\ref{DFR}) as the shaded squares
 in Fig.~\ref{fig:BosonFRP}. The corresponding loop evaluation gives
\bea
 G_{\underline{8}}(Q_G) \, 
\label{G2(QD)} \; = 
\;   \, - \frac{C_G}{8 \pi^2} \, \frac{1}{24}
\, \langle \frac{\alpha_s}{\pi}G^2 \rangle \, .
\eea
For the gluon condensate in (\ref{G2(QD)}) we take the 
value
$\langle \frac{\alpha_s}{\pi}G G \rangle^{1/4} = 310$ 
MeV, in agreement, within uncertainties, with
 the lattice results and values used in refs. 
\cite{Bertolini:1997nf,BeFE00,HE}.

Both expressions (\ref{G2(Q6)}) and (\ref{G2(QD)}) are concrete examples of 
  the general separation of SD and LD effects, 
contained in the factorized form of the $G_j$'s
  in eq. (\ref{LMes})
for our procedure of bosonisation.
By performing the steps above, we have 
all ingredients that are necessary to
estimate the new contribution of our dipole penguin operator $Q_G$
on the same  footing as 
 the previously calculated contributions.

\section{Results and Discussion}

In the present Letter we are investigating a new dipole penguin operator
 that is the off-shell partner of the standard chromomagnetic
 penguin dipole operator $Q_{11}$. Namely,
in addition to the chromomagnetic penguin dipole operator (\ref{11}) 
phrased \cite{BuBL96} as the mass insertions on external quark lines,
there is, as explained in eqs. (\ref{eq:lagg}) and (\ref{eq:lags}),
an additional dipole operator $Q_G$ displayed in eq. (\ref{Q_G})
 corresponding to
 off-shell momenta for external quarks confined in hadrons.
Such Lamb-shift like effects in strong interactions
 are in another context very recently discussed in ref. \cite{BrodShr}.

 The new dipole penguin operator $Q_G$ studied here has
 several attractive features.
We have shown that it has the same bosonisation as the
standard $Q_6$ operator. Accordingly, $Q_G$ dominates over $Q_{11}$
which is higher order in chiral expansion.
 Indeed, the bosonised form ${\cal{L}}^{(4)}(Q_{11})$ results in the
 suppressed $K \rightarrow 2 \pi$ amplitude 
${\cal A}(K^0 \rightarrow \pi^+ \pi^- ; Q_{11})$ presented in eq. (\ref{2.13}).
 Thereby, the bosonised 
form ${\cal{L}}^{(2)}(Q_{6})$ leads to the referent
 $K \rightarrow 2 \pi$ amplitude
 ${\cal A}( K^0 \rightarrow \pi^+ \pi^- ;\, Q_6)$  in eq. (\ref{A2(Q6)}), 
and the two operators, $Q_G$ and $Q_6$, differ only in their respective coefficients,
  $G_{\underline{8}}(Q_G)$  in eq. (\ref{G2(QD)}) and
 $G_{\underline{8}}^{(2)}(Q_6)$  in eq. (\ref{G2(Q6)}).

The ratio between $Q_G$ and $Q_6$
contributions can now be read from the LD hadronic factors and the 
SD Wilson coefficients contained in the coefficients $G_{\underline{8}}(Q_G)$ 
 and $G_{\underline{8}}^{(2)}(Q_6)$ :
\bea 
\rho  \; \equiv \; \frac{{\cal A}( K^0 \rightarrow \pi^+ \pi^- ;\,
  Q_G)}{{\cal A}( K^0 \rightarrow \pi^+ \pi^- ;\, Q_6)} \; = \;
\frac{C_G}{C_6} \, h \; \; ,
\label{rho}
\eea
where the hadronic factor $h$ denotes the ratio of the respective LD pieces, 
\bea
h \; = \; \frac{f_\pi^2 \, \langle \frac{\alpha_s}{\pi}G^2 \rangle}{24
  \cdot 8 \pi^2 \cdot 16 \, L_5 \, |\langle \bar{q} q  \rangle |^2 }
\; \; .
\eea
By substituting the numerical values, including
$\langle \frac{\alpha_s}{\pi}G G \rangle^{1/4} = 310$ MeV
and  $L_5 \simeq 1.4 \times 10^{-3}$, we obtain  
the hadronic factor $h \simeq 0.011$.

Finally, by employing the appropriate Wilson coefficients 
for the  operators $Q_G$ and $Q_6$ given in Table 1, we
obtain for the CP-violating and the CP-conserving parts of the ratio in eq. (\ref{rho}):
\begin{eqnarray}
{\rho_{\text{CP-violating}}} & = &  \frac{y_G}{y_6}  \, h \;   \, \simeq \;   \, 0.05 \;   ,
\end{eqnarray}
\begin{eqnarray}
{\rho_{\text{CP-conserving}}} & =&  \frac{z_G}{z_6}  \, h \;   \, \simeq \;   \, 0.04 \;  .
\label{CPV:CPC}
\end{eqnarray}
This represents $\sim 5$ \% of the referent $Q_6$ contribution, to which
 it adds
both in CP-violating and CP-conserving parts. In particular, there is a net coherent 
contribution from the CP-violating off-shell amplitudes,  the one from
 the new dipole operator considered here, and  the previously calculated
 off-shell self energy contribution of  $\sim$ 15 \% to $\epe$ 
in \cite{Bergan:1996sb}. In conclusion, within the chiral quark model 
approach, we obtain in total an increase of
$\sim$ 20 \% with respect to the leading $Q_6$ contribution to
the CP-violating ratio $\epe$ from off-shell operators.
This result is still within the uncertainty of the theoretical value of
Trieste-Oslo group, and although slightly higher than the world average 
 (\ref{eps}), it is
 closer to the new preliminary KTeV result  
$\epe=(19.2 \pm 1.1 \pm 1.8) \cdot 10^{-4}$ \cite{NewKTeV}.

\subsubsection*{Note added}
After the first submission of this Letter we noticed another Letter \cite{BrodShr} addressing 
the effects of Lamb shift type  in QCD.
We also became aware of the  new KTeV result
based on doubling of the statistics
and an improved control of systematics 
\cite{NewKTeV}. 

\subsubsection*{Acknowledgement}
This work was supported by the Ministry  of Science, Education and Sport of the
Republic of Croatia under contract No. 119-0982930-1016.
J.~O.~E. thanks the Physics Department of University in Zagreb for its
hospitality, and 
K.~K. gratefully acknowledges
the hospitality of the Department of Physics in Oslo.
J.~O.~E.  is supported in part by the Norwegian
research council and  by the European Commission RTN
network, Contract No. MRTN-CT-2006-035482 
(FLAVIAnet).

\end{document}